\documentclass[12pt]{article}
\usepackage[utf8]{inputenc}
\usepackage[T1]{fontenc}
\usepackage{graphicx}
\usepackage{epstopdf} 
\usepackage{color}
\input{epsf}
\usepackage{amsmath}
\usepackage{amsfonts}
\usepackage{amstext}
\usepackage{amssymb}
\usepackage{revsymb}
\usepackage{mathrsfs}
\usepackage{accents}
\newcommand{\be}{\begin{equation}}
\newcommand{\ee}{\end{equation}}
\newcommand{\bear}{\begin{eqnarray}}
\newcommand{\ear}{\end{eqnarray}}

\begin{document}
\title{Atwood's machine with a massive string}
\author{Nivaldo A. Lemos\\
\small
{\it Instituto de F\'{\i}sica - Universidade Federal Fluminense}\\
\small
{\it Av. Litor\^anea, S/N, Boa Viagem, Niter\'oi,
24210-340, Rio de Janeiro - Brazil}\\
\small
{\it nivaldo@if.uff.br}}

\date{\today}

\maketitle
\begin{abstract}
	
The dynamics of Atwood's machine with a string of significant mass are described by the Lagrangian formalism, providing an eloquent example of how the Lagrangian approach is a great deal simpler and so much more expedient than the Newtonian treatment.  

\end{abstract}


Either with a massless or a massive pulley,  Atwood's machine \cite{Greenslade} is a standard example for the application of the laws of Newtonian  mechanics. In its most idealized manifestation,  the pulley and the string are supposed to be massless and the pulley is assumed to be
mounted on a frictionless axle. The mass of the pulley is easily taken into account by the dynamics of the pulley's rotational motion about its fixed axis \cite{Serway}. In these two cases treated in the textbooks the string is taken  to be massless. Both the idealized and the realistic case, in which account is taken of the masses of the string and the pulley as well as of the friction in the pulley's bearings, illustrate the principles involved in the application of Newton's laws  \cite{Kofsky,Martell}. Atwood's machine is a multipurpose mechanical system, which allows  one to investigate from Stokes's law \cite{Greenwood} to variable-mass rocket motion \cite{Greenwood2}.

\begin{figure}[h!] 
	\centering
	\includegraphics[width=0.3\linewidth]{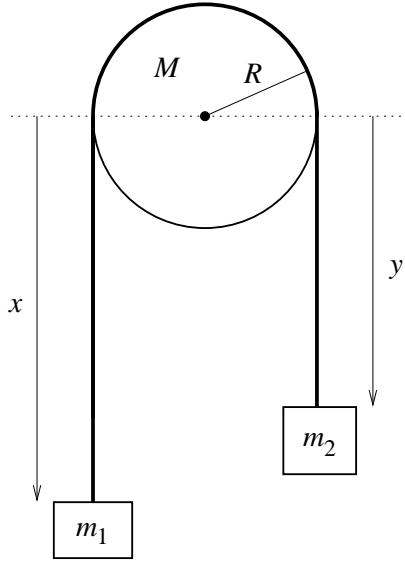}
	\caption[Circle]{Two masses attached to a massive string that goes around a pulley.}
\label{masses-pulley}
\end{figure}

Here we wish to determine the effect of the mass of the string on the motion of  Atwood's machine by means of the Lagrangian formalism.  Consider the Atwood's machine depicted in Fig. \ref{masses-pulley}, in which  the string has uniform linear mass density $\lambda$. The mass $m_s$ of the string is not supposed to be negligible in comparison with $m_1$ and $m_2$. We assume that  the string is inextensible and does not slide on the pulley, which requires enough static friction between the string and the pulley.
On the other hand, we assume that the pulley is mounted on a frictionless axle.

The treatment of this problem by Newtonian mechanics is cumbersome, since it requires the consideration of free-body  diagrams for the masses $m_1$ and $m_2$, 
the pulley together with the segment of the string in contact with it, and the hanging parts of the string at each side of the pulley. In order to find the acceleration of  $m_1$, for instance, Newton's second law for each separate body must be written down and the four tensions at the ends of the hanging pieces of the string have to be eliminated by algebraic manipulation of the equations \cite{Tarnopolski}.

We tackle the problem by means of Lagrangian dynamics instead. We have the constraint
\begin{equation}
\label{constraint}
x+ y = \ell 
\end{equation}
where $\ell$ is the length of the hanging part of the string, that is, the total length of the string minus the length of its segment that touches the pulley. 
Since the string is inextensible and does not slide on the pulley, the masses $m_1$, $m_2$ and all points of the string move at the  same speed
$v = \vert {\dot x} \vert$, while the angular speed of the pulley is $\omega = v/R$. Thus, the kinetic energy of the system is 
\begin{equation}
\label{kinetic}
T = \frac{m_1}{2}v^2 + \frac{m_2}{2}v^2 + \frac{m_s}{2}v^2 + \frac{I}{2}\omega^2\, = \frac{1}{2} (m_1 + m_2 + m_s+ I/R^2) {\dot x}^2  \,  ,
\end{equation}	
where $m_s$ is the mass of the string and $I$ is the pulley's moment of inertia with respect to the rotation axis. For instance, if the pulley is a homogeneous disk then $I=MR^2/2$ 

Taking the  horizontal plane that contains the pulley's axle as the  plane of zero gravitational potential  energy, the potential energy of the system is the sum of the gravitational potential energies of $m_1$, $m_2$, and of the hanging pieces of the string at each side of the pulley. Thus 
\begin{equation}
\label{potential}
V = -m_1gx - m_2gy - \lambda  x g\frac{x}{2} - \lambda  y g\frac{y}{2} \, ,
\end{equation}
where we have used  the fact that the gravitational potential  energy of an extended body is determined by the position of its centre of mass. An immaterial constant has been dropped, namely the gravitational potential energy of the non-hanging part of the string.
With the use of the constraint (\ref{constraint}) the potential energy becomes
\begin{equation}
\label{potential2}
V = -(m_1 - m_2)gx - \frac{\lambda  g}{2} x^2 -  \frac{\lambda  g}{2} (\ell - x)^2  \, ,
\end{equation}
where the irrelevant additive constant $-m_2g\ell$ has been discarded.
 
Thus, the Lagrangian is  
\begin{equation}
\label{Lagrangian}
L = T - V  =  \frac{1}{2} (m_1 + m_2 + m_s+ I/R^2) {\dot x}^2  + (m_1 - m_2)gx + \frac{\lambda  g}{2} x^2 + \frac{\lambda  g}{2} (\ell - x)^2 \,  .
\end{equation}
Lagrange's equation
\begin{equation}
\label{Lagrange-equation}
  \frac{d}{dt} \bigg(\frac{\partial L}{\partial {\dot x}} \bigg) - \frac{\partial L}{\partial x} = 0
\end{equation}
yields at once
\begin{equation}
\label{equation-of-motion}
 (m_1 + m_2 + m_s+ I/R^2) {\ddot x}  = (m_1 - m_2)g +  2\lambda g x - \lambda g \ell \, .
\end{equation}
This equation predicts that $\, {\ddot x} > 0\, $ if $\, m_1 + \lambda  x > m_2 + \lambda  (\ell - x)$, that is, if the total mass on the left of the pulley is larger than the total mass on the right, which is the correct physical condition for the mass $m_1$ to  accelerate downward. Furthermore, in the limit of a massless string ($\lambda =0$) we recover the constant acceleration
\begin{equation}
\label{massless-limit}
  {\ddot x}  = \frac{m_1 - m_2}{m_1 + m_2 +  I/R^2}\, g  
\end{equation}
that is found in the textbooks. One can hardly fail to appreciate that the Lagrangian approach to this problem is substantially simpler and much more expedient than  the traditional Newtonian treatment \cite{Tarnopolski}.

When the string is massive the acceleration is not constant. The equation of motion (\ref{equation-of-motion}) takes the form
\begin{equation}
\label{differential-equation-of-motion}
 {\ddot x} -  b^2 x = a \, ,
\end{equation}
where
\begin{equation}
\label{a-and-b}
  a = \frac{(m_1 - m_2- \lambda  \ell )g}{m_1 + m_2 + m_s+ I/R^2} \, , \,\,\,\,\,\,\,\,\,\,   b = \bigg[ \frac{2 \lambda g}{m_1 + m_2 + m_s+ I/R^2}\bigg]^{1/2} \, .
\end{equation}
The general solution to the inhomogeneous linear differential equation (\ref{differential-equation-of-motion}) is the sum of a particular solution with the general solution to the homogeneous equation, namely
\begin{equation}
\label{general-solution}
  x (t) = -\frac{a}{b^2} + A\cosh bt + B\sinh bt 
\end{equation}
where $A$ and $B$ are arbitrary constants, with
\begin{equation}
\label{a-over-b2}
 \frac{a}{b^2}= \frac{m_1 - m_2- \lambda  \ell}{2\lambda} \, .
\end{equation}

The natural initial conditions  $x(0)= x_0$, ${\dot x}(0) =0$ lead to 
\begin{equation}
\label{constants-natural-initial-condition}
 A= x_0 + \frac{a}{b^2} \, , \,\,\,\,\,\,\,\,\,\, B=0\, ,
\end{equation}
whence
\begin{equation}
\label{solution-natural-initial conditions}
  x (t) = -\frac{a}{b^2} +  \bigg( x_0 + \frac{a}{b^2} \bigg) \cosh bt \, . 
\end{equation}

A particularly simple case is $m_1=m_2=0$. Then the string moves owing only to  the unbalanced weights of its hanging parts at each side of the pulley.   In this case
\begin{equation}
\label{a-and-b-special}
 \frac{a}{b^2} = - \frac{\ell}{2}  \, , \,\,\,\,\,\,\,\,\,\,  b = \bigg[\frac{2 \lambda g}{m_s+I/R^2}\bigg]^{1/2} ,
\end{equation}
and it follows that
\begin{equation}
\label{solution-initial conditions-special}
  x (t) = \frac{\ell}{2} +  \Bigl(x_0 - \frac{\ell}{2}\Bigr) \cosh bt \,. 
\end{equation}
As physically expected, if $x_0=\ell /2$ then $x(t) \equiv \ell /2$, that is, the string remains at rest in unstable equilibrium. 

\begin{figure}[h!] 
	\centering
	\includegraphics[width=0.45\linewidth]{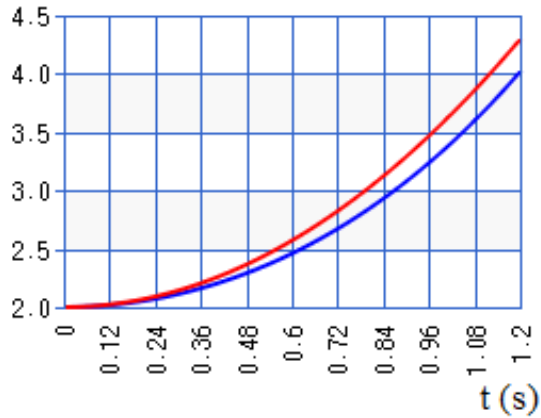}
	\caption[Circle]{Position, in metres, of $m_1$  as a function of time, in seconds,  for a massless string (upper line) and for a massive string (lower line) for the values of the parameters and initial conditions given in the text.}
\label{comparison-graph}
\end{figure}

The time scale of the motion is set by the time constant $\tau = 1/b$. 
For $t$ bigger than a few time constants the common speed of the masses and the string  grows exponentially.
This exponential behavior is due to the fact that as  the string falls to one side   
more mass is transferred to that side, increasing its weight unbalance with respect 
to the other side. Suppose $m_1=0.20$ kg, $m_2=0.10$ kg and consider a rope with $\lambda = 0.02$ kg/m and  $\ell = 4.0$ m. For  a 
typical 
laboratory pulley we have  $I= 1.8 \times 10^{-6}$ kg m$^2$ and $R=2.5$ cm, from which 
it follows that $m_s = \lambda (\ell + \pi R ) = 0.084$ kg. From equations   (\ref{a-over-b2}) and (\ref{a-and-b})  we find 
\begin{equation}
\label{a-over-b2-and-b-numerical}
  \frac{a}{b^2} = 0.50\,\mbox{m} \, , \,\,\,\,\,\,\,\,\,\, b= 1.0 \, \mbox{s}^{-1} \, .
\end{equation}
 Therefore, if $x_0 = 2.0$ m equation (\ref{solution-natural-initial conditions}) yields
\begin{equation}
\label{solution-initial-conditions-numerical}
 x (t) = -0.50 + 2.5 \, \cosh t \, , 
\end{equation}
with $x$ in metres and $t$ in seconds. If the string's mass were neglected, the mass $m_1$ would fall with the constant acceleration ${\tilde a}= 3.2$ m/s$^2$ given by equation (\ref{massless-limit}). With the same initial conditions, the instantaneous position of $m_1$ would be given by
\begin{equation}
\label{solution-initial-conditions-massless-string-numerical}
 {\tilde x} (t) = 2.0 + 1.6\, t^2  \, . 
\end{equation}
Figure \ref{comparison-graph} shows a comparison graph of ${\tilde x} (t)$ (upper line) and $x(t)$ (lower line) from $t=0$ to $t=1.2$ s, when $x$ reaches its largest physically allowed value, namely $x=4.0$ m. In the present case the  effect of the string's mass is that of slowing down the motion. If the string's mass is disregarded then  $x=4.0$ m is reached at ${\tilde t}=1.1$ s instead of the correct $t=1.2$
s, an 8\%  error that certainly requires a careful experimental set up  to be  detected. If $x_0$ is smaller the discrepancy is larger and more easily detected: for $x_0 = 0.50$ m one has $t=2.2$ s whereas $\tilde t = 1.5$ s.

\end{document}